\documentclass[amsmath,amssymb, aps, prl, twocolumn, notitlepage]{revtex4-1}

\usepackage{graphicx}
\usepackage{dcolumn}
\usepackage{bm}
\usepackage[usenames]{color}
\usepackage{slashed}
\usepackage[utf8]{inputenc}
\usepackage{amsfonts}
\usepackage{amsmath}
\usepackage{amssymb}
\usepackage{hyperref}
\usepackage{latexsym}
\usepackage{color}

\begin{document}
\title{Quantum-limit Hall effect with large carrier density in topological semimetals}
\author{Guang Yang, Yi Zhang}
\email{frankzhangyi@gmail.com}
\affiliation{International Center for Quantum Materials, School of Physics, Peking University, Beijing, 100871, China}

\begin{abstract}
The quantum-limit Hall effect at $\nu = nh/eB\sim O(1)$ that hosts a variety of exotic quantum phenomena requires demanding strong magnetic field $B$ and low carrier density $n$. We propose to realize quantum-limit Hall effect even in the presence of large carrier density residues $n_e$ and $n_h$ relative to the magnetic field $B$ in topological semimetals, where a single Fermi surface contour allow both electron-type and hole-type carriers and approaches charge neutrality as $n_e\sim n_h$. The underlying filling factor $\nu = |n_e-n_h|h/eB$ explicitly violates the Onsager's relation for quantum oscillations. 
\end{abstract}

\maketitle
 
 A two-dimensional electron system may display the integer quantum Hall effect with exactly quantized Hall conductance $\sigma_{xy} = \nu e^2/h$ when subjected to a strong magnetic field \cite{IQHE1980,TKNN1982,Laughlin1981IQH}, where the filling factor $\nu$ takes integer values. In the quantum limit $\nu = nh/eB \sim O(1)$, the discrete quantization of the lowest Landau levels becomes essential, making ground for exotic quantum phenomena such as the fractional quantum Hall effect \cite{TsuiFQH,LaughlinFQH, Read1990,Haldane1983FQH,Jain1989FQH,ZSC1989FQH}, the chiral anomaly \cite{Nielsen1983,Ishikawa1984CA,Alekseev1998CA,2016Signatures}, etc. However, strong magnetic fields $B$ and low carrier densities $n$ are among the prerequisites to suppress the geometric phases for the quantum limit, restricting its experimental realizations and widespread applications. 

\begin{figure}
\includegraphics[width=.95\linewidth]{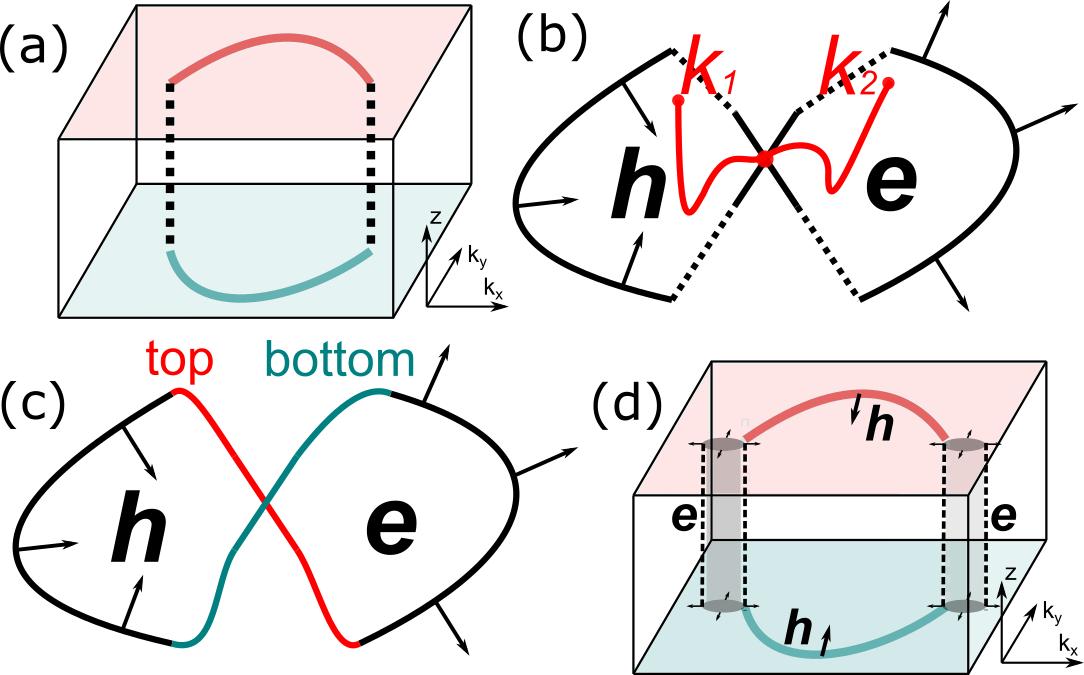}
\caption{(a) The Weyl orbit in a topological semimetal consists of the Fermi arcs on the surfaces and the chiral Landau levels in the bulk (black dashed lines). (b) The energy $E(k_2)<\mu$ ($E(k_1)>\mu$) inside an electron (hole) pocket suggests that at least one point between $k_1$ and $k_2$ is on the Fermi surface $E=\mu$. Even in this best-case scenario, the contour may further divide into separate pockets up a magnetic field or inter-band coupling. (c) The Fermi arcs on topological semimetal top and bottom surfaces are stable and protected by spatial separation. (d) The topological semimetal may have hole-type Fermi arcs and electron-type bulk Fermi surfaces, or vice versa. The black arrows denote Fermi velocity directions.}
\label{fig:elec+hole}
\end{figure}

 Semi-classically, the charge carriers in a magnetic field move along constant-energy field-normal cyclotron orbits, whose overall phase's Bohr-Sommerfeld quantization $\phi = 2\pi \nu + \gamma$ offers an intuitive perspective for the discrete Landau levels \cite{Onsager1952, Lifshitz1956,QianNiu2010}. As the magnetic field changes, the highest occupied Landau level follows, and hence a variety of properties oscillate periodically in $1/B$. The frequency of quantum oscillations obeys the Onsager relation and is proportional to the field-normal Fermi surface area, thus providing a useful experimental probe to the material electronic structure and considered a robust Fermi liquid behavior \cite{Patrick2009onsager,Baym:1320624}. More recently, the cyclotron orbit extended its concept to the Weyl orbit \cite{Potter2014, frank2016, Moll2016, Frank2019weylorbitknot} in topological semimetal \cite{Wan2011, Xu2015,Ding2015WSM,Ashvin2018RMP,WHM2015WSM,Liu864,FZDX2013DSM}, which consists of the Fermi arcs on the surfaces and the chiral ($0^{th}$) Landau levels in the bulk (Fig. \ref{fig:elec+hole}a) and leads to quantum Hall effect in three spatial dimensions \cite{Zhang2017, Wang2017, Zhang2018,LHL20203DQHE}.

 We note that electron pocket and hole pocket contribute opposite geometric phases to their cyclotron orbits and their carrier densities compensate each other, therefore a partially electron-type and partially hole-type Fermi surface can suppress its overall filling factor $\nu = \left|n_e - n_h\right|h/eB$, especially when $n_e \sim n_h$. Conventionally, it seems unlikely for a pocket to be electron-type and hole-type at the same time, see Fig. \ref{fig:elec+hole}b \footnote{An example is the Fermi surface at a Lifshitz transition \cite{Lifshitz1960}.}. Here, however, we demonstrate a simple counterexample illustrated schematically in Fig. \ref{fig:elec+hole}c, where the crossing between the electron-type region and hole-type region of a Weyl orbit in topological semimetal slab consists of spatially-protected Fermi arcs on opposite surfaces. Further, we discuss the experimentally-accessible scenario where the surface Fermi arcs and the bulk Weyl fermions have opposite carrier types, as in Fig. \ref{fig:elec+hole}d. In both cases, the system can approach the quantum limit $\nu\sim O(1)$ despite large carrier density residues $n_e$ and $n_h$ in relative comparison with the strength of magnetic field $B$.

 \emph{Model and method.}\textemdash For concreteness, let's consider the following three-dimensional Weyl semimetal tight-binding model with slab geometry of thickness $L_z$ \cite{PhysRevB.86.195102}:
\begin{eqnarray}
H_0&=&\sum_{z=1}^{L_z}\sum_{k_\parallel}\varepsilon_{k_\parallel,z}c_{k_\parallel,z}^{\dagger}c_{k_\parallel,z}^{\vphantom\dagger}
+\left(h_{k_\parallel,z}c_{k_\parallel,z+1}^{\dagger}c_{k_\parallel,z}^{\vphantom\dagger}+\mbox{h.c.}\right) \nonumber\\
& &\varepsilon_{k_\parallel,z}=\left(2t \cos k_x + 2t \cos k_y - \epsilon_0\right)\times\left(-1\right)^{z-1} \label{eq:Ham}\\
& &h_{k_\parallel,z}=t_z+2t'\sin(k_y) \times\left(-1\right)^{z-1} \nonumber
\end{eqnarray}
 where $k_\parallel = \left(k_x, k_y\right)$ is the in-plane momentum. 
 Without loss of generality, we set $t=1.0$, $t_z=1.0$, and $t'=0.5$ unless noted otherwise. 
 Hole-type Fermi arcs appear on the top layer $z=L_z$ and bottom layers $z=1$, and center around $k_\parallel = (0,0)$ following the condition $\varepsilon(k_\parallel)=\mu$. On the other hand, the bulk states are characterized by the momentum-space Hamiltonian:
\begin{equation}
H_{0}(\vec{k})=\varepsilon(k_\parallel)\sigma^z+2t_z\cos(k_z/2)\sigma^x + 4t'\sin(k_z/2)\sin(k_y)\sigma^y
\end{equation} 
 where $k_z$ is the momentum in the $\hat z$ direction, and $\sigma$'s are Pauli matrices with the first (second) row and column describing the odd (even) layers. The two bulk bands $\pm E(\vec{k})$ meet at zero energy only at the two Weyl nodes $(\pm k_W,0,\pi)$, $k_W=\cos^{-1}(\epsilon_0/2t-1)$. In the bulk, the chiral Landau levels at the Weyl nodes connect the Fermi arcs on both surfaces, and together, they form a closed contour, namely, the Weyl orbit (Fig.\ref{fig:elec+hole}a).
 
 In the presence of an external magnetic field $\vec{B}$, we employ the Landau gauge $\vec{A}=\left(0, \Phi_z x - \Phi_x z, -\Phi_y x\right)$, where $\Phi_{d}$ is the magnetic flux quantum $\Phi_0=h/e$ per plaquette perpendicular to the $\hat d$ direction, $d = x, y, z$. Since $k_y$ remains a good quantum number, we can calculate the properties of this Hamiltonian efficiently with the recursive Green's function method, such as the density of states (DOS) $\rho(\mu)=-\frac{1}{\pi L_x L_z}\sum_{x,z}\text{Im} G(x,z;x,z;\mu+i\delta)$ at the chemical potential $\mu$, where the imaginary part $\delta=0.001$ gives a small level broadening. 

\begin{figure}
\includegraphics[width=.49\linewidth]{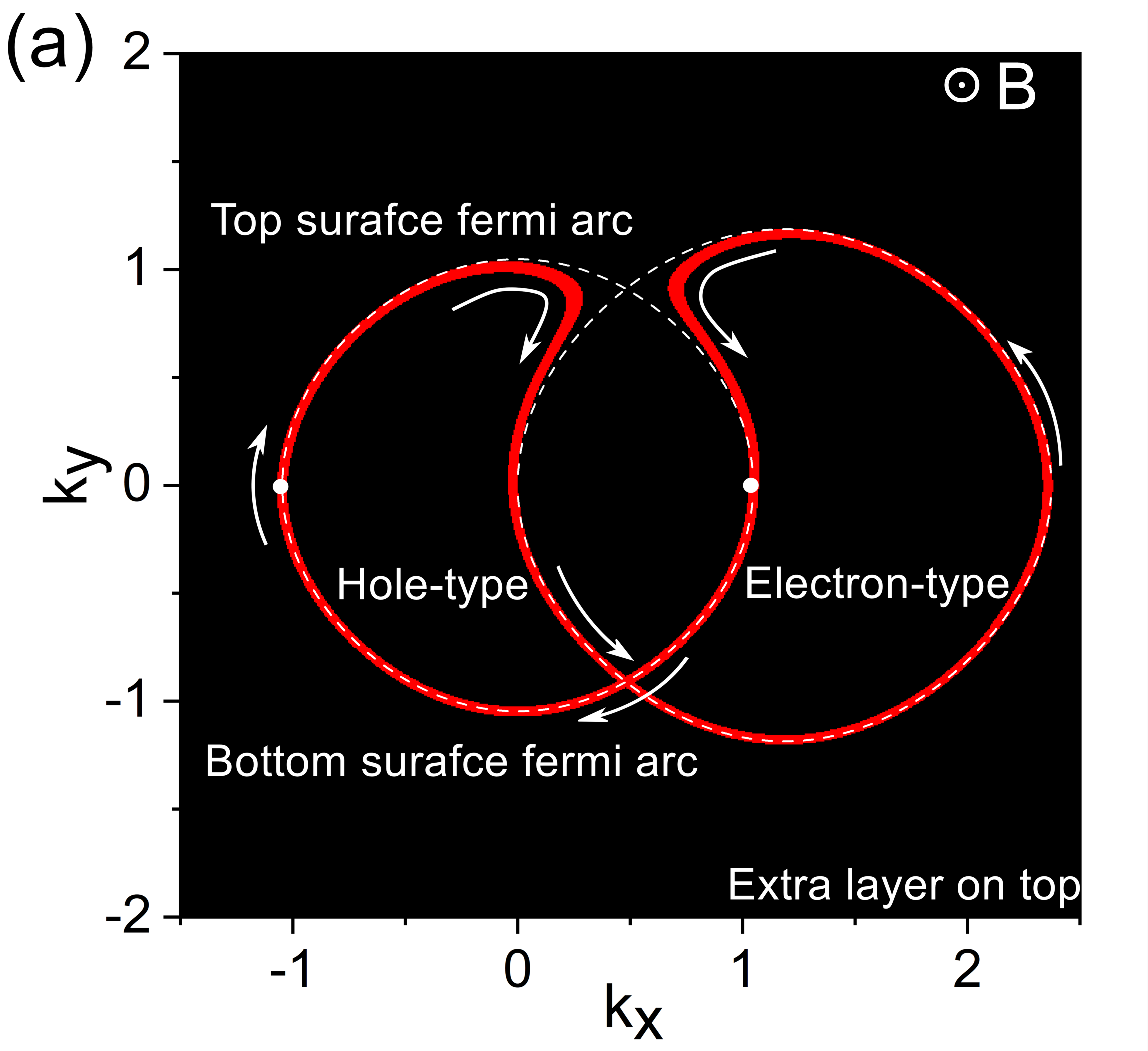}
\includegraphics[width=.49\linewidth]{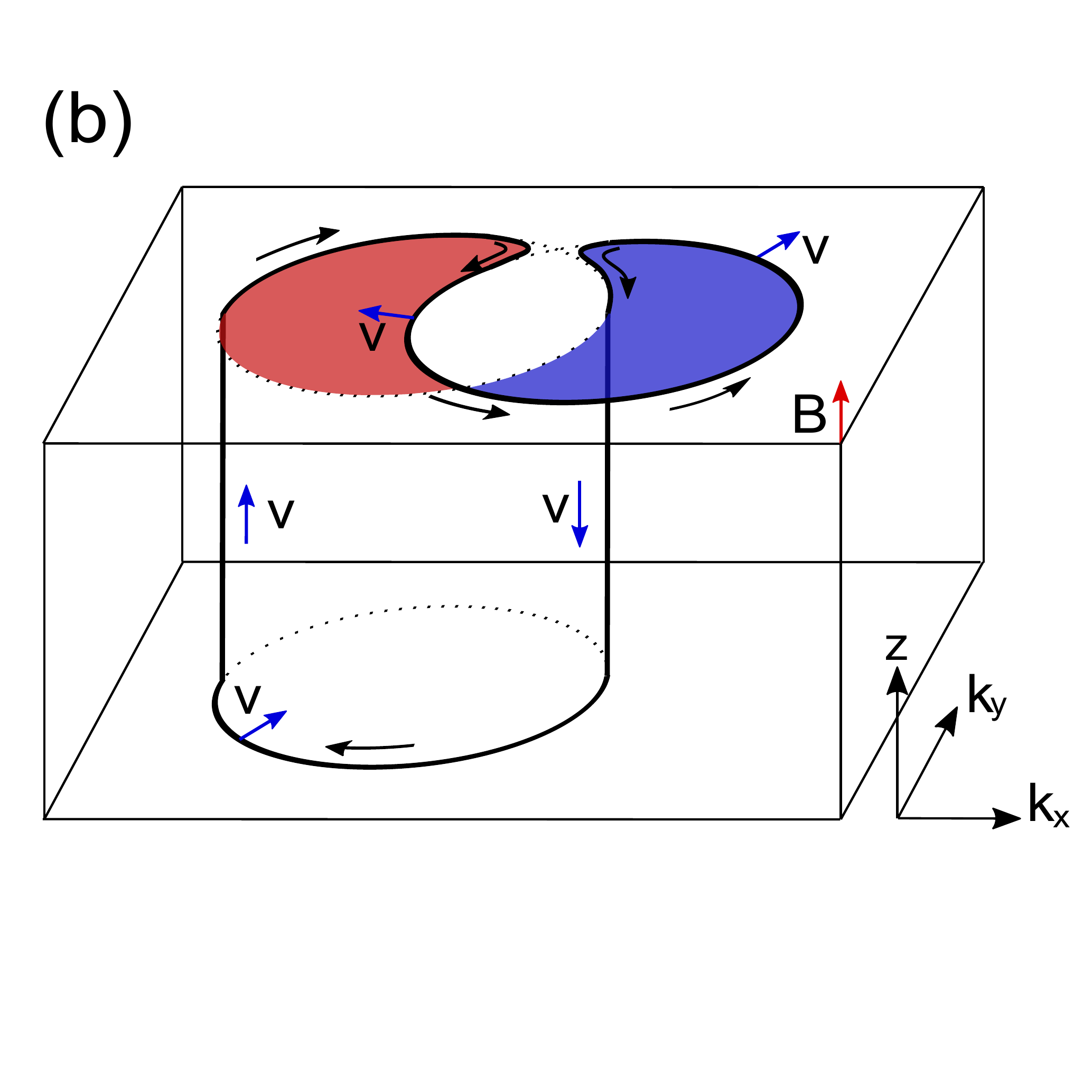}
\includegraphics[width=.9\linewidth]{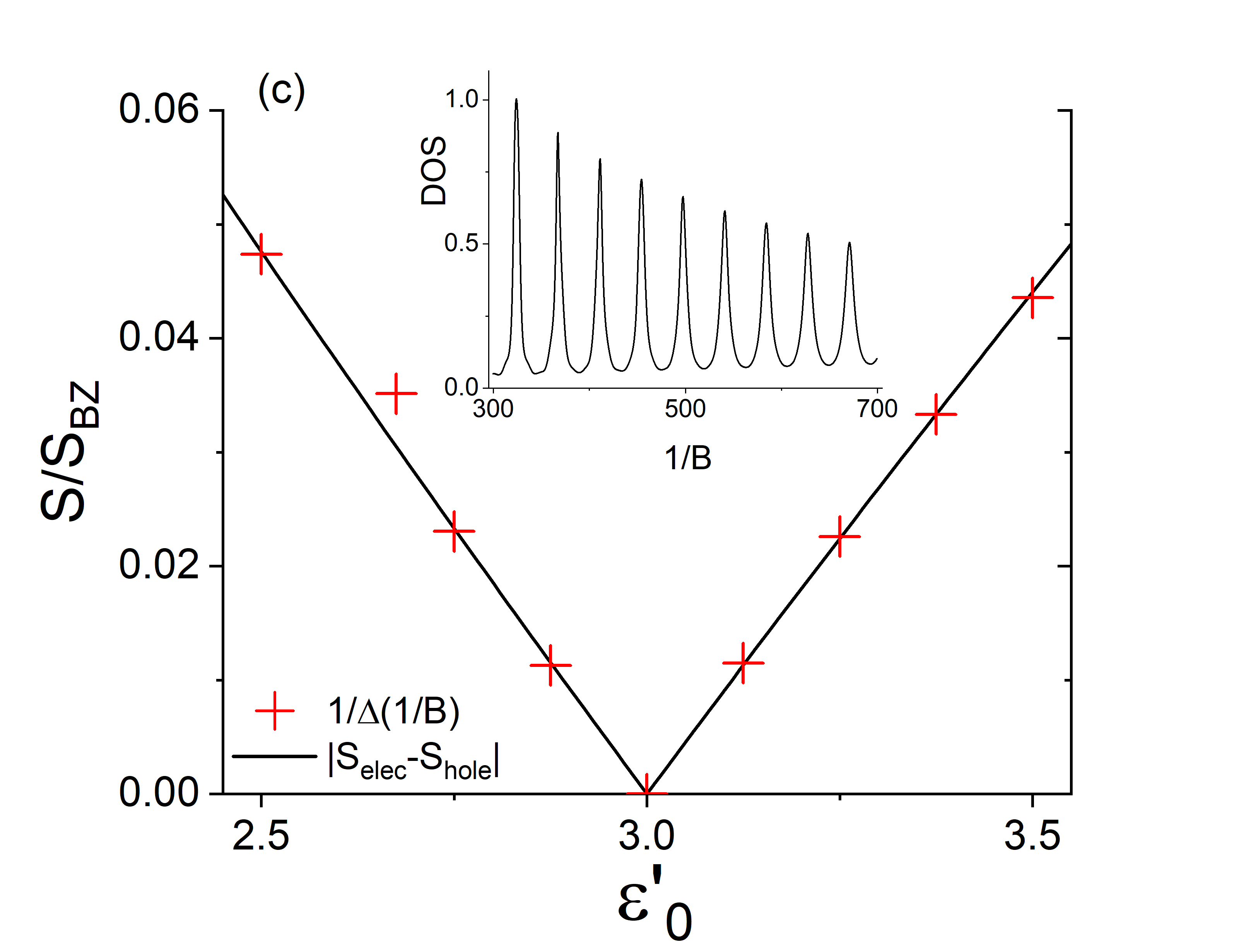}
\caption{(a) The red curve denotes the reconstructed Fermi surface contour in the $k_x-k_y$ plane for $H=H_0+H_1$. The white dots are locations of the bulk Weyl nodes where the electrons tunnel between the surfaces.
The white arrows are the directions electron move semi-classically along the contour. $\epsilon_0=3.0$, $\epsilon'_0=2.75$. (b) A schematic diagram of the Weyl orbit for $H=H_0+H_1$ in slab geometry. The red and blue region represents the hole-type pocket and the electron-type pocket, respectively, which now coexist in a single contour. (c) The periods of the quantum oscillations $\Delta(1/B)$ versus $\epsilon'_0$ is consistent with the absolute difference between the area $S_{elec}$ of the electron-type region and the area $S_{hole}$ of the hole-type region in (a). 
The inset illustrates the quantum oscillations in the range $1/B\in [300,700]$ for $\epsilon_0=3.0$, $\epsilon'_0=2.75$.}
\label{fig:weylorbit}
\end{figure}

 \emph{Heuristic surface-state example.}\textemdash We first illustrate the scenario in Fig. \ref{fig:elec+hole}c by including an extra layer $H=H_0 + H_1$ and altering the Fermi arc on the top surface:
\begin{eqnarray}
H_1 &=& \sum_{k_\parallel}\varepsilon'_{k_\parallel}c_{k_\parallel,L_z+1}^{\dagger}c_{k_\parallel,L_z+1} + \Delta \left(c_{k_\parallel,L_z+1}^{\dagger}c_{k_\parallel,L_z}  + \mbox{h.c.}\right) \nonumber \\ 
& &\varepsilon'_{k_\parallel}= -2t \cos \left(k_x-k_0 \right) - 2t \cos k_y + \epsilon'_0
\label{eq:toplayer}
\end{eqnarray}
 The dispersion $\epsilon'_{k_\parallel}$ gives an electron pocket centered at $k_\parallel=(k_0,0)$, which interacts with the top-surface Fermi arc upon a coupling $\Delta=0.25$. The Fermi arc on the bottom surface is spatially separated and remains intact \cite{Frank2019weylorbitknot}. As shown in Fig. \ref{fig:weylorbit}a, the reconstructed Fermi surface on the $k_\parallel$ plane encloses both an electron-type region and a hole-type region. We set $\epsilon_0=3.0$, $L_z=65$, $\mu=0$ at the energy of the Weyl node, and $k_0=\cos^{-1}(\epsilon'_0/2t-1)$.

 As we vary the magnetic field $B$ in the $\hat z$ direction, we observe clear signatures of quantum oscillations in $\rho(\mu)$ corresponding to the Weyl orbit, see Fig. \ref{fig:weylorbit}c inset. Interestingly, the frequency of quantum oscillations $f=1/\Delta(1/B)=\left|S_{elec}-S_{hole}\right|/S_{BZ}$ is consistent with the \emph{absolute difference} between the area of the electron-type region $S_{elec}$ and the area of the hole pocket $S_{hole}$ (Fig. \ref{fig:weylorbit}b), instead of the total area $S_{hole}+S_{elec}$ enclosed by the Fermi surface. $S_{BZ}$ is the area of the Brillouin zone. For instance, when $\epsilon'_0=2.75$ and $\epsilon_0=3.0$, $S_{elec}=7.30\%S_{BZ}$, $S_{hole}=4.92\%S_{BZ}$, whose sum $S_{elec}+S_{hole}$ is more than four times larger than their difference $S_{elec}-S_{hole}$. Consequently, this example constitutes an explicit violation of the Onsager relation. Semi-classically, the electron moves counter-clockwise (clockwise) around the electron-type (hole-type) region, see Fig. \ref{fig:weylorbit}a. Therefore, while proportional to their respective areas, the contribution of these two segments to the overall geometric phase differs in sign.

\begin{figure}
\includegraphics[width=.98\linewidth]{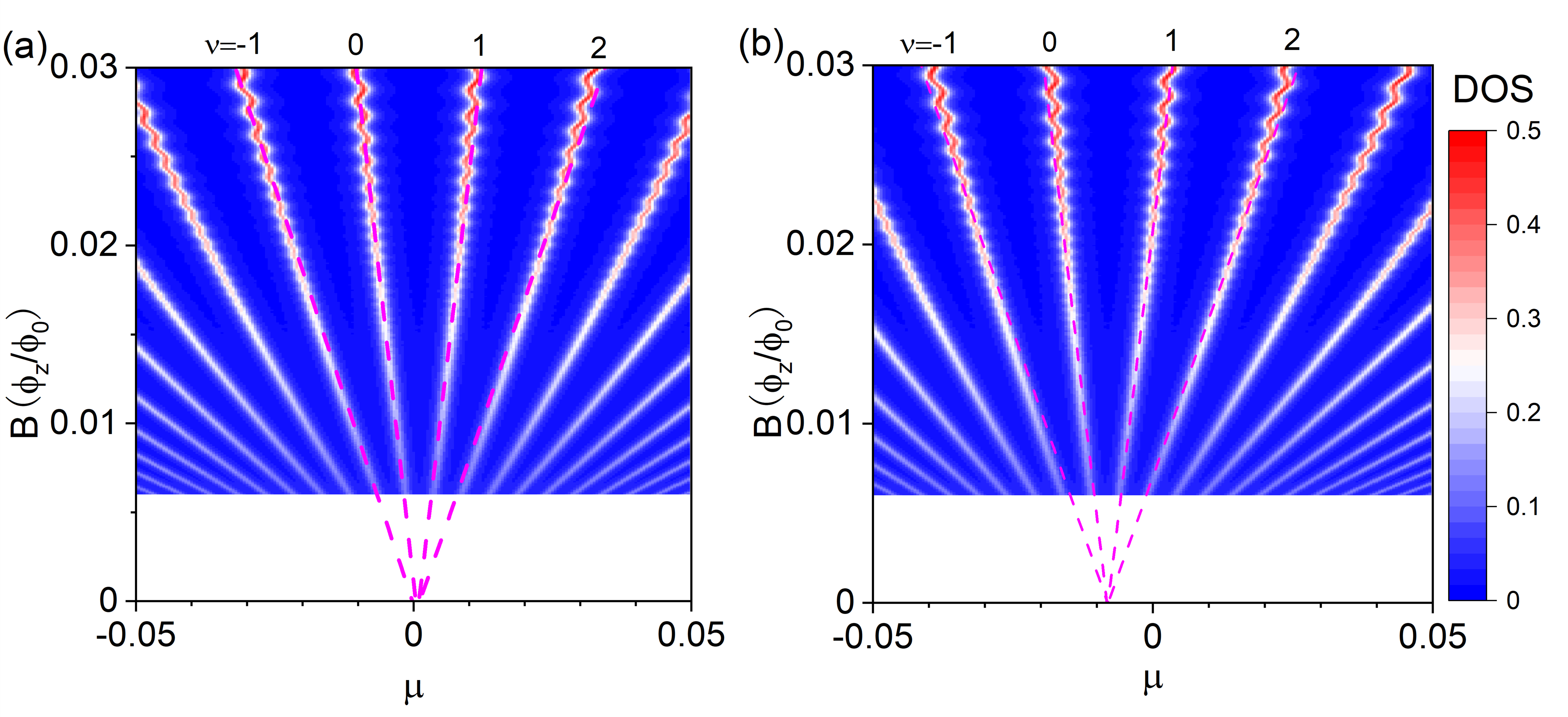}
\caption{The Landau fan diagrams depict the dependence of the lowest Landau levels of the model $H=H_0+H_1$ in Eqs. \ref{eq:Ham} and \ref{eq:toplayer} on the magnetic field $B$ and the chemical potential $\mu$ at (a) $\epsilon'_0=3.0$ and (b) $\epsilon'_0=2.98$. The color scale represents the DOS $\rho(\mu)$, and the magenta dashed lines are linear guides to the eye.}
\label{fig:fan}
\end{figure}

 Such cancellation effect between the electron carriers and hole carriers is also in effect for the quantum Hall effects in general, such as the conditions for the quantum limit $\nu=\left|n_e-n_h\right|h/eB$ at relatively larger magnetic fields. The Landau fan diagrams in Fig. \ref{fig:fan} demonstrate the entrance of the quantum limit. Importantly, as the filling factor is dominated by the absolute difference between the electron density $n_{e}$ and the hole density $n_{h}$, the necessary magnetic field $B$ to reach a target $\nu$ can be suppressed by orders of magnitude, especially when the electron and hole densities are close $n_{e} \sim n_{h}$. For example, the quantum limit occupying only the first landau level is approached at $B(\Phi_z/\Phi_0)=0.0048$ for $H=H_0+H_1$ with $\epsilon_0'=3.02$ and $\mu=0.0$, while the required magnetic field is $B(\Phi_z/\Phi_0)=0.199$ for a fully hole-type system $H = H_0$ with similar settings to reach the quantum limit - the cancellation effect leads to small effective carrier density despite large carrier density residues and reduced the required magnetic field by over 97\%. Therefore, cyclotron orbits with both electron-type and hole-type carriers is indeed physically possible and can potentially allow quantum-limit Hall effect despite large carrier density residues, as we have showcased with the Weyl orbit in topological semimetal. 

\begin{figure}
\centering
\includegraphics[width=.49\linewidth]{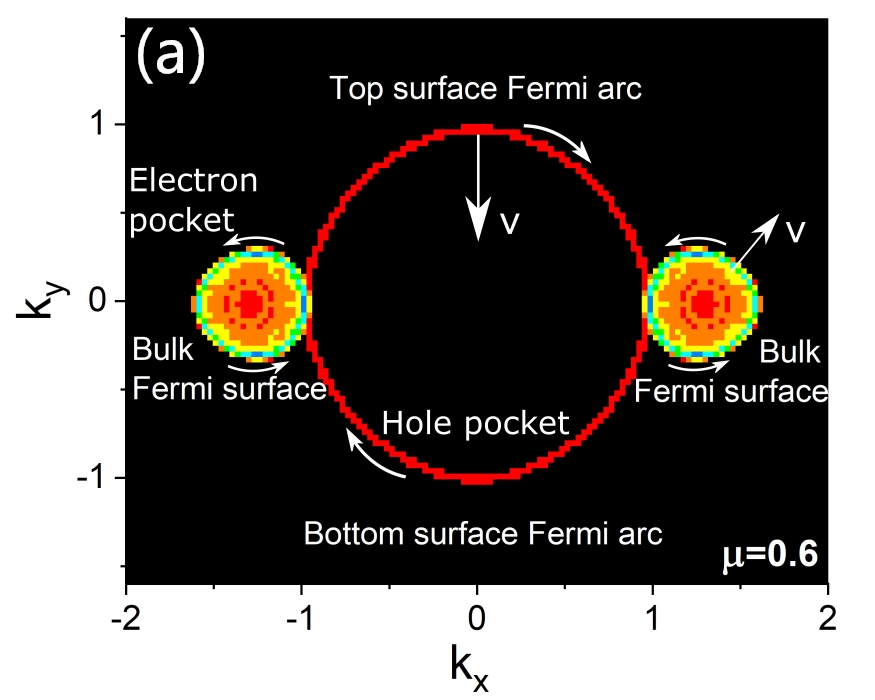}
\includegraphics[width=.49\linewidth]{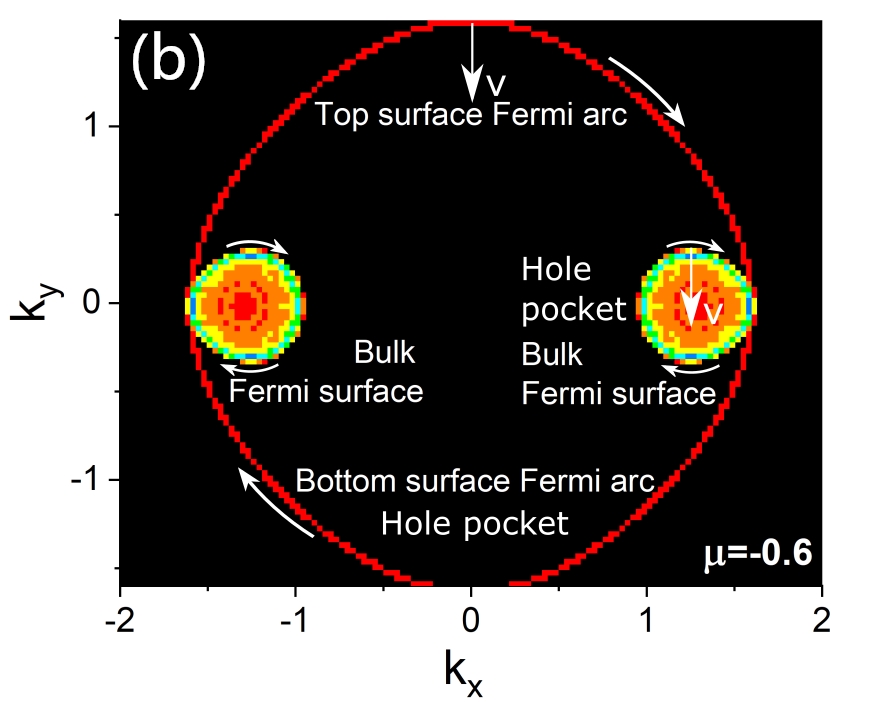}
\includegraphics[width=.49\linewidth]{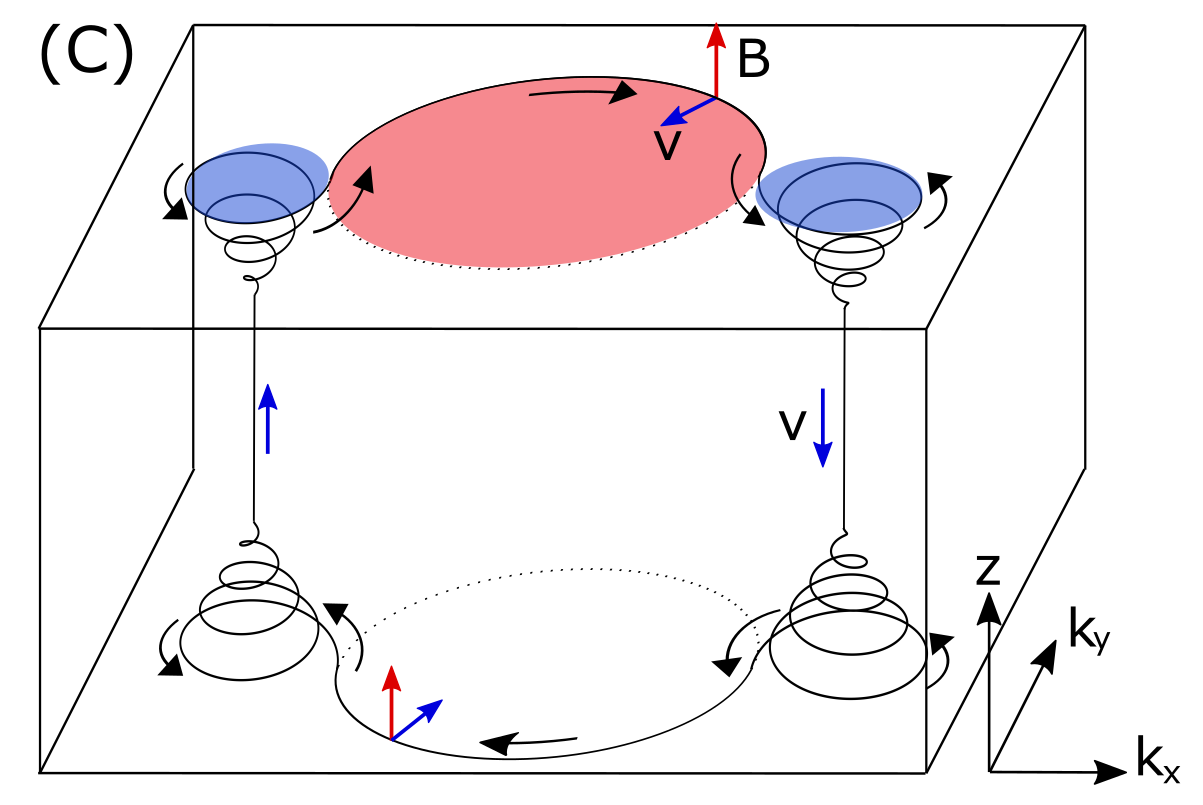}
\includegraphics[width=.49\linewidth]{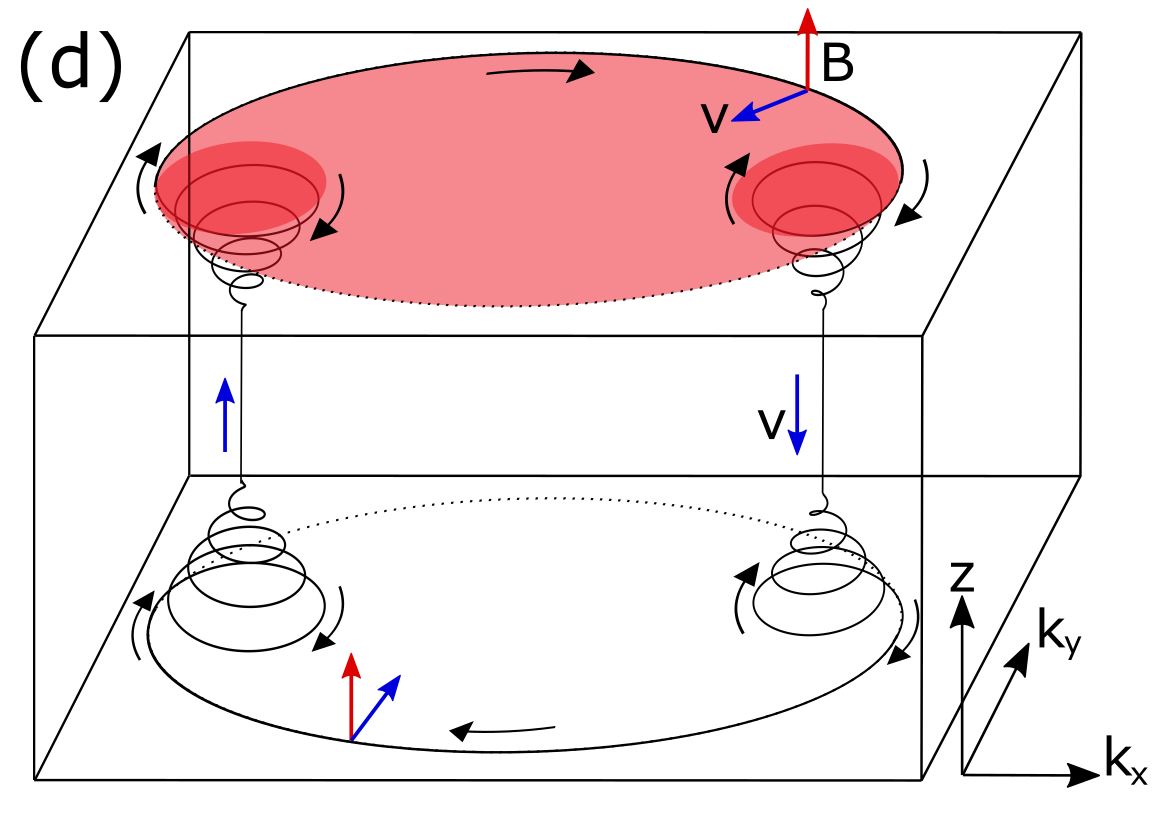}
\caption{The contours of surface Fermi arcs and bulk Fermi surfaces in the $k_x-k_y$ plane at Fermi energy (a) $\mu=0.6$ and (b) $\mu=-0.6$, respectively. $\epsilon_0=2.5$, and $L_z=145$. The Fermi arcs are hole-type in both cases, whereas the bulk states are two electron pockets or two hole pockets around the Weyl nodes, respectively. The directions of the electron semiclassical motion and Fermi velocities are also shown. (c) and (d) are schematic plots of the Weyl orbit in the presence of a magnetic field, which consists of Fermi arcs on the surfaces, the chiral Landau levels in the bulk, and the transition regions in between, where the directions of the semiclassical motions of the charge carriers are (c) opposite and (d) identical to those along the Fermi arcs, respectively.}
\label{fig:qos}
\end{figure}

 \emph{Practical example with surface-bulk dichotomy.}\textemdash Next, we discuss the scenario in Fig. \ref{fig:elec+hole}d where the different carrier types amongst the Weyl orbit descend from material surface and bulk, respectively. Let's consider the original Weyl semimetal model $H_0$ in Eq. \ref{eq:Ham} at a Fermi energy sufficiently away from the Weyl nodes. While the surface Fermi arcs amount to a hole-type pocket, the bulk states are two electron (hole) pockets around the Weyl nodes for $\mu>0$ ($\mu<0$), see examples of the Fermi surface contours in the $k_{\parallel}$ plane in Fig. \ref{fig:qos}. The area enclosed by the Fermi arcs is $S_{FA} = 7.71\% S_{BZ}$ ($S_{FA}=20.04\% S_{BZ}$) at $\epsilon_0=2.5$ and $\mu=0.6$ ($\mu=-0.6$), and $0.76\% S_{BZ}$ for each of the bulk Fermi surfaces. On the other hand, the DOS $\rho(\mu)$ shows clear signatures of quantum oscillations in a magnetic field $B$ in the $\hat z$ direction that correspond to the Weyl orbits, yet their periods at $\Delta(1/B)=16.77\pm 0.25$ for $\mu=0.6$ and $\Delta(1/B)=3.89\pm 0.03$ for $\mu=-0.6$ are inconsistent with the expectations ($\Delta(1/B)=12.97$ at $\mu=0.6$ and $\Delta(1/B)=4.99$ at $\mu=-0.6$) from the Onsager relation by non-negligible margins.

 To understand the discrepancies, we revisit the Weyl orbit components for Fermi energy away from the Weyl nodes. Previous studies have established that the components localized on the surface and deep inside the bulk are the Fermi arcs and the chiral Landau levels descending from the Weyl nodes, respectively \cite{Potter2014,frank2016}. However, intermediate transitions necessarily exist for a nonzero $\mu$ around the four turning points, where the electron ventures a finite separation between the the Fermi arc' endpoint and the $k_\parallel$ of the Weyl node, while its wave packet gradually evolves from the surface Fermi arc states localized in the real space $z=1, L_z$ to the chiral Landau level states localized in the momentum space $k_z=\pm k_F$, or vice versa \footnote{We discuss more details on the property of the transition regions in Supplementary Materials, and work on a consistent, quantitative semiclassical theory that also includes the transitions is in preparation.}. The winding directions at the transitions depend on the carrier type of the bulk states and are opposite (identical) to that of the hole-type Fermi arcs at $\mu=0.6$ ($\mu=-0.6$), see Figs.\ref{fig:qos}c and \ref{fig:qos}d for illustration, enclosing magnetic flux and accumulating Berry phase during the process, therefore contributing an opposite (additional) geometric phase to the overall quantization condition:
\begin{eqnarray}
\frac{h}{e}\frac{S_{T}-S_{FA}}{B_z}+ \gamma_T + \left(2k_F+2k_W\sin\theta\right) L_z \sec \theta = 2\pi \nu
\nonumber \\
\label{eq:3dfit}
\end{eqnarray}
where $k_F$ is the Fermi vector of the chiral Landau levels, $\theta$ is the magnetic field's tilting angle from $\hat z$ to $\hat x$ direction, and $2k_W\sin\theta=\left(\vec{k}_{W1}-\vec{k}_{W2}\right)\cdot \hat B$ is the displacement between the Weyl nodes projected along the magnetic field. The contributions $-hS_{FA}/eB_z$ and $\left(2k_F+2k_W\sin\theta\right) L_z \sec \theta $ to the geometric phase are from the surface Fermi arcs and the bulk chiral Landau levels, respectively \cite{Potter2014, frank2016}, where we have adopted the convention that contributions of electron-type (hole-type) sections are positive (negative). In addition, we note that the four transition regions along the Weyl orbit also contribute to the geometric phase and give rise to the violation of the Onsager relation. We attribute their contributions $hS_{T}/eB_z +\gamma_T$ to two origins: (1) an orbital piece $hS_{T}/eB_z$ as the trajectory encloses the magnetic flux, sometimes multiple times, and (2) a Berry phase $\gamma_T$ as the (even-odd layer) pseudo-spin rotates during the transitions. In particular, the effective momentum-space area $S_T>0$ ($S_T<0$) for $\mu=0.6$ ($\mu=-0.6$) negates (adds to) $S_{FA}$, therefore increasing (decreasing) the period of quantum oscillations. 


\begin{figure}
\centering
\includegraphics[width=.95\linewidth]{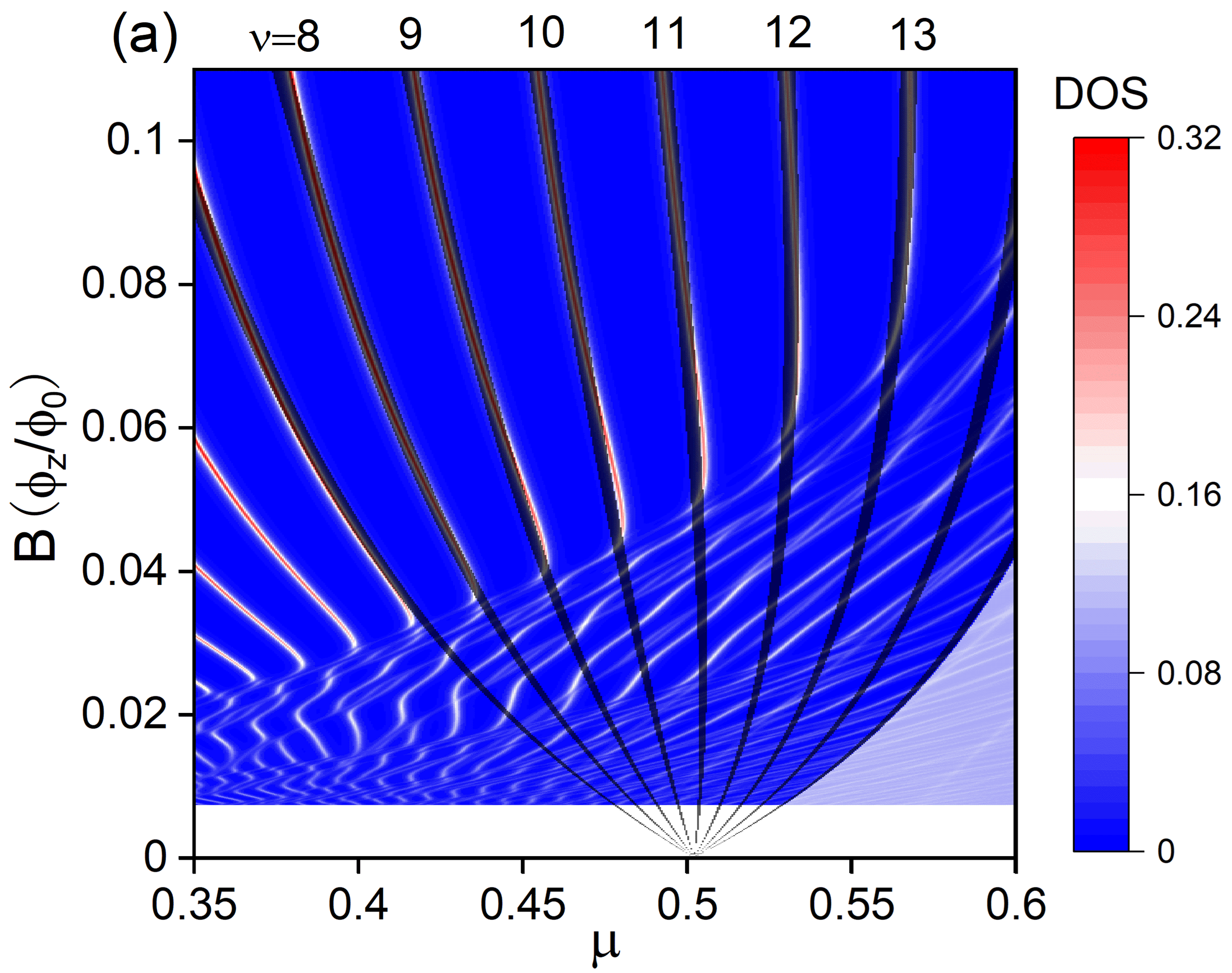}
\includegraphics[width=.95\linewidth]{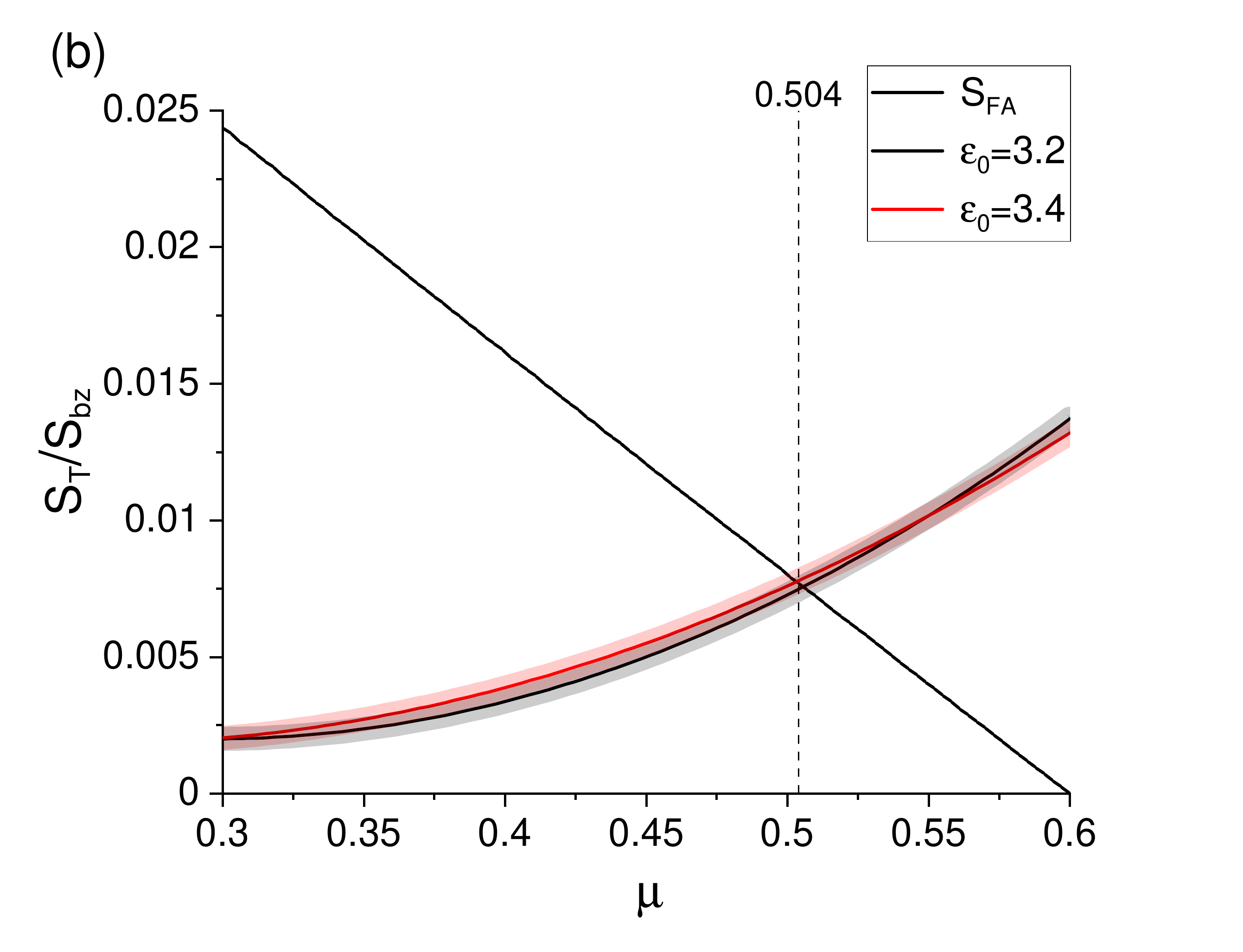}
\caption{(a) The Landau fan diagram shows the dependence of the lowest Landau levels of the model in Eq. \ref{eq:Ham} on the magnetic field $B_z$ and the Fermi energy $\mu$ sufficiently above the Weyl node. $\epsilon_0=3.4$, $L_z=145$, and $\theta=0$. The color scale represents the DOS $\rho(\mu)$, and sharp Landau levels are visible. The slope changes sign at $\mu \sim 0.51$ suggesting the full cancellation of $S_{FA}$ and $S_T$, and the $B_z$ dependence vanishes. The blur on the bottom right is due to the hybridization with the conventional bulk Landau levels, which kick in at $\mu \propto \sqrt{B}$. The black lines are Eq. \ref{eq:3dfit} with polynomial fits to the transition region's contributions $hS_{T}/eB_z +\gamma_T$. (b) The effective momentum-space area of the transition regions $S_{T}$ extracted from the fit increase rapidly as the Fermi energy moves further away from the Weyl node and surpasses $S_{FA}$ - the hole-pocket area of the Fermi arcs - at around $\mu \sim 0.50$.}
\label{fig:fan2}
\end{figure}

 The Landau fan diagram in Fig. \ref{fig:fan2}a at relatively larger magnetic fields indicates quantum-limit Hall effect behaviors despite the large carrier densities both on the surface and in the bulk at Fermi energy $\mu\sim 0.5$. Thankfully, the electron carrier density $n_{elec}$ of the surface states proportional to $S_{FA}$, and the hole carrier density $n_{hole}$ of the transition regions proportional to $S_{T}$, cancel due to their opposite carrier types within the same Fermi surface contour. We set $\epsilon_0=3.4$, $L_z=145$ and $\theta=0$ in these calculations. We note that both $S_{T}$ and $\gamma_T$ may depend explicitly on the transition trajectory and thus the Fermi energy $\mu$ and the magnetic field $B_z$, and perform a simple polynomial fit to $hS_{T}/e +\gamma_T B_z$ with all other quantities in Eq. \ref{eq:3dfit} derived analytically. More details on the model fitting are available in Supplemental Materials. The effective $B_z$-independent momentum-space area $S_T$ from the transition region are illustrated in Fig. \ref{fig:fan2}b. $S_T$ increases rapidly as a function of the Fermi energy $\mu$, especially above $\mu \sim 0.3$, consistent with the fact that both the $k_{\parallel}$ cross-section and the extents of winding of the transitions increase with $\mu$. Therefore, even a moderate transition region can significantly impact the overall geometric phase as long as the Fermi energy is sufficiently far away from the Weyl nodes. 


 We note that while the fan diagram in Fig. \ref{fig:fan2} exhibits a near cancellation of the $B_z$-dependent surface ($S_{FA}$) and transition ($S_{T}$) contributions, there remains a non-negligible constant geometric phase $2k_F L_z$ from the bulk chiral Landau levels at finite $\mu$. To make our `quantum limit' live up to its name, we can tilt the magnetic field away from $\hat z$ to a so-called magnetic angle \cite{frank2016}, so that $\left(\vec{k_{W1}}-\vec{k_{W2}}\right)\cdot \hat B = -2k_F$ and the thickness contribution $(2k_W\sin\theta+2k_F)L_z sec \theta $ vanishes. We include detailed results and analysis for a tilted magnetic field to Supplemental Materials. 

 The above proposal relies upon the proper tuning of the Fermi energy and the magnetic field angle and is generally more applicable for experiments and realizable with existing Weyl semimetals. The separation between the Weyl nodes should be appropriate: a large separation may lead to Fermi arcs and surface carrier density too large for a proper chemical potential in the bulk to compensate; a small separate may lead to the pair-wise annihilation of Weyl nodes in a magnetic field \cite{patrick2017,JiashuangTaP2017,Ramshaw2018}. It further helps the material not to possess irrelevant pockets in the explored energy window to avoid potential intervention. For instance, the Weyl semimetal material HfCuP was predicted to possess a hole-type surface Fermi arcs, and electron-type bulk states only around the Weyl nodes \cite{MENG2020WSM}. A signature of the cancellation of the geometric phase is the sign change of $n_e{\sim}n_h$ thus the Hall number $n_H=B/e\rho_{xy}$ upon tuning or gating, while properties concerning the overall charge carrier density $n_e+n_h$, e.g. the longitudinal conductivity $\sigma = (n_e+n_h) e^2 \tau /m$, remain little changed. Our discussions also generalize straightforwardly to Dirac semimetals, which possess Weyl orbits below a threshold value of the magnetic field \cite{Potter2014}. 


 \emph{Conclusion.}\textemdash We discover that electron-type and hole-type carriers may coexist in the same Fermi surface contour and contribute oppositely to the geometric phase and quantum Hall phenomena, which can approach the quantum limit $\nu=|n_e-n_h|h/eB$ with a much smaller magnetic field than commonly needed with properly-tuned despite large carrier density residues $n_e{\sim}n_h$. We illustrate our conclusions with the Weyl orbit in topological semimetals and note that the physics also holds for quantum systems where a single cyclotron orbit traverses different segments with opposite carrier types, such as surface states and bulk states. 

 \emph{Acknowledgements.}\textemdash We thank Ryuichi Shindou for helpful discussions and comments. G.Y. and Y.Z. are supported by the start-up grant at Peking University. The computation was supported by High-Performance Computing Platform of Peking University.
 
\bibliography{refs}

\end{document}